\documentclass[amsmath,notitlepage,amssymb,aps,prx,superscriptaddress,twocolumn,footinbib]{revtex4-1}

\usepackage{amsmath}
\numberwithin{equation}{section}
\renewcommand{\theequation}{\arabic{section}.\arabic{equation}}

\usepackage{braket} 

\usepackage{dsfont} 

\usepackage{lipsum} 

\usepackage{float}

\usepackage{graphicx}
\usepackage{dcolumn}
\usepackage{bm}

\usepackage[colorlinks]{hyperref}
\hypersetup{
    citecolor=[rgb]{0,0,1},
    anchorcolor=[rgb]{1,0,1}, 
    linkcolor=[rgb]{0,0,1},   
    urlcolor=[rgb]{0,0,0.9},
    linktocpage={true}}
\usepackage{wasysym}
\usepackage{mathtools} 

\usepackage[export]{adjustbox} 

\usepackage[mathscr]{euscript} 

\usepackage{diagbox}

\newcommand{\bk}{{\bf k}}
\newcommand{\bK}{{\bf K}}

\newcommand{\bQ}{{\bf Q}}

\begin{document}

\title{Translation symmetry-enriched toric code insulator}

\author{Pok Man Tam}
\affiliation{Department of Physics and Astronomy, University of Pennsylvania, Philadelphia, PA 19104, USA}

\author{J\"orn W. F. Venderbos}
\affiliation{Department of Physics, Drexel University, Philadelphia, PA 19104, USA}%
\affiliation{Department of Materials Science \& Engineering, Drexel University, Philadelphia, PA 19104, USA}%

\author{Charles L. Kane}
\affiliation{Department of Physics and Astronomy, University of Pennsylvania, Philadelphia, PA 19104, USA}

\begin{abstract}
We introduce a two-dimensional electronic insulator that possesses a toric code topological order enriched by translation symmetry. This state can be realized from disordering a weak topological superconductor by double-vortex condensation. It is termed the toric code insulator, whose anyonic excitations consist of a charge-$e$ chargon, a neutral fermion and \textit{two} types of visons. There are two types of visons because they have constrained motion as a consequence of the fractional Josephson effect of one-dimensional topological superconductor. Importantly, these two types of visons are related by a discrete translation symmetry and have a mutual semionic braiding statistics, leading to a symmetry-enrichment akin to the type in Wen's plaquette model and Kitaev's honeycomb model. We construct this state using a three-fluid coupled-wire model, and analyze the anyon spectrum and braiding statistics in detail to unveil the nature of symmetry-enrichment due to translation. We also discuss potential material realizations and present a band-theoretic understanding of the state, fitting it into a general framework for studying fractionalizaton in strongly-interacting weak topological phases.

\end{abstract}

\maketitle

\section{\label{sec1}Introduction}

Over the past decades, symmetry and topology have emerged as two central and interwoven organizing principles in the study of condensed matter physics. In the case of weakly interacting systems, symmetry-protected topological (SPT) phases---a class which includes topological insulators (TIs) and superconductors (TSCs)---have been predicted theoretically and in a number of cases discovered experimentally \cite{HasanKane2010, QiZhang2011}. Internal symmetries, such as time-reversal and particle-hole symmetries, give rise to so-called strong SPT phases with protected gapless boundary modes \cite{Ryu2010, Chiu2016}. In addition to strong SPT phases, there exist \textit{weak} SPT phases, which can be viewed and constructed as stacks of strong SPT phases from a lower dimension. Importantly, the distinction of weak SPTs from a trivial phase requires an additional discrete translation symmetry along the stacking direction, which prevents hybridization of pairs of stacked layers. Prototypical examples of weak SPTs include the three-dimensional (3D) weak TI \cite{3DTI2007, MooreBalents2007, Roy2009}, which harbors an even number of surface Dirac modes, and the two-dimensional (2D) weak TSC \cite{TeoHughes2013, weakTSCstability2014}, which harbors a pair of counter-propagating Majorana edge modes. While the basic properties of weak topological phases are well understood in the weakly-interacting regime, less is known about the effect of strong interactions, which may lead to exotic correlated phases and phenomena either on the boundary or in the bulk. The purpose of this paper is to study the effect of strong interactions on weak topological phases, and in particular to address the interplay between translation symmetry and topology. To this end, we focus on the paradigmatic example of a strongly interacting weak TSC in two dimensions. 

Strong interactions can give rise to correlated quantum phases with emergent fractionalized quasiparticles known as anyons. Such quantum phases are referred to as topological orders \cite{WenTO1990, Wen2004QFT}. Well-known examples of topological order include fractional quantum Hall states and quantum spin liquid states \cite{Savary2016, Hansson2017CFTReview}. In the case of the former, recent experimental evidence for the fractional statistics of Laughlin quasiparticles has been reported \cite{ENS2020, Purdue2020}. Perhaps the simplest example of topological order, however, is of the $\mathbb{Z}_2$ type, which was first studied in the context of frustrated quantum antiferromagnets \cite{Anderson1987, Kivelson1987, ReadSachdev1991, Wen1991} and later reconsidered in the form of Kitaev's toric code toy model, as well as Wen's plaquette model \cite{Kitaev2003TQC, Wenplaquette2003}. Given the compelling appeal of the toric code model, in this paper we refer to the concerned $\mathbb{Z}_2$ topological order as the toric code topological order. It features four types of anyons: $\mathds{1}$, $e$, $m$ and $f=e \times m$, where $e$ and $m$ are self-bosons which obey a mutual $\pi$-braiding (semionic) statistics, as well as the $\mathbb{Z}_2$ fusion rule: $e^2 = m^2 =\mathds{1}$. Remarkably, as pointed out by Hansson \textit{et al.} \cite{Hansson2004SCisTO}, toric code topological order not only arises in spin systems, but also in conventional superconductors, where $f$ is interpreted as the Bogoliubov fermion and $m$ as the superconducting vortex. It is then natural to wonder to what extent and how the structure of the topological order is modified in unconventional superconductors. More specifically, it is natural to ask whether a 2D weak TSC can provide a platform for a toric code topological order \textit{enriched} by translation symmetry. Here we answer this question in the affirmative by constructing an insulator from a strongly interacting weak TSC and showing that the resulting anyon spectrum exhibits symmetry-enrichment.

In general, a symmetry-enriched topological (SET) order can exhibit many interesting properties in addition to the fusion and braiding properties of anyons~\cite{Ran2013SET, Hung2013SET, Lu2016SET, Cheng2016translationSET,Chen2017SET, Cheng2019SET}. For instance, it can feature fractionalized symmetry quantum numbers, such as the electric charge of Laughlin quasiparticles. In this paper, we focus on another aspect: the \textit{permutation} of anyon types by a symmetry transformation. This has been famously demonstrated in spin systems such as Wen's plaquette model and Kitaev's honeycomb model \cite{Wenplaquette2003, Kitaev2006exactly}, in which $e$-particles transform into $m$-particles (and vice versa) under a discrete translation. This phenomenon is sometimes known as ``weak symmetry breaking", where the pattern of quasiparticles breaks the symmetry of the Hamiltonian as well as the ground state of the system \cite{Kitaev2006exactly, Rao2020}. In the spinless electronic context, such a translation SET order has been recently proposed in what is termed the ``nondiagonal" quantum Hall state, which possesses a charge sector of a Laughlin state and a neutral sector characterized by a $\mathbb{Z}_p$ toric code, with the latter featuring a weak breaking of translation symmetry \cite{Tam2021}. 


In this work, we use a coupled wire model approach and introduce a 2D electronic insulator that realizes a $\mathbb{Z}_2$ toric code topological order enriched by translation symmetry. Unlike the nondiagonal quantum Hall state, this does not require a quantizing magnetic field and does not produce an extra charge sector. Here we find that the discrete translation by one wire permutes the $e$-particle with the $m$-particle of the toric code, which is similar to observations made in previous works in the context of spin models. We refer to this topological order as the ``symmetry-enriched toric code insulator", and we will argue that it can arise from a \textit{disordered} 2D weak TSC, through a competition between charge-density wave instability and superconducting instability that leads to double-vortex condensation. 

Before presenting our microscopic model in the next section, we begin by providing an intuitive understanding of the symmetry-enriched toric code insulator, based on the physics of fractional Josephson effect in topological superconductors. 



\subsection{The 2D weak topological superconductor}

\begin{figure}[t]
   \includegraphics[width=\columnwidth]{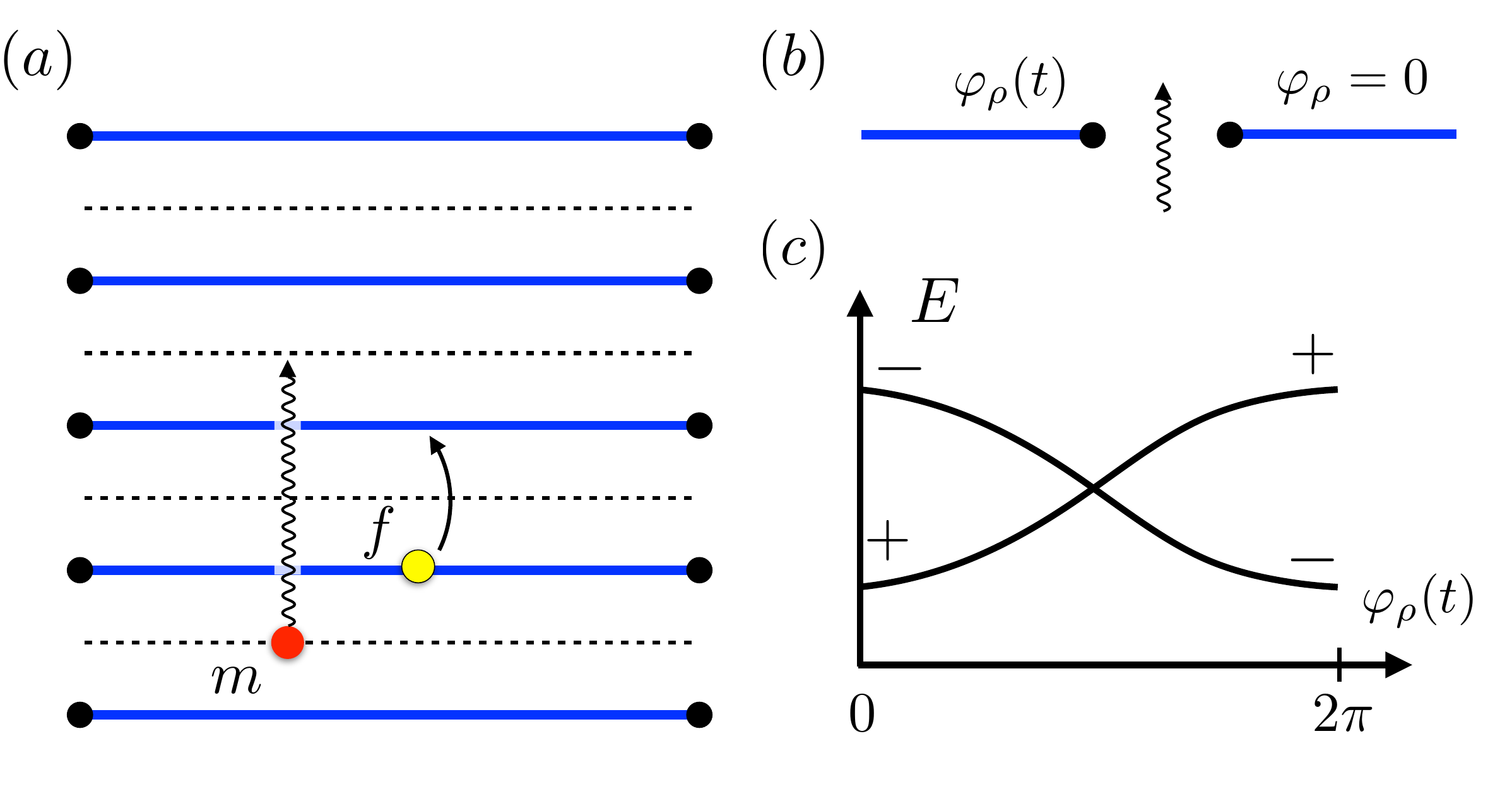}\centering
  \caption{\small{(a) shows the wire model of a weak TSC, where each wire represents a 1D TSC (and black dots represent Majorana end modes). Two types of point-like excitations in the bulk are depicted: the $h/2e$ vortex $(m)$ lives on links (dashed lines), and the Bogoliubov fermion $(f)$ lives on wires (blue lines). (b,c) illustrate the fractional Josephson effect: tunneling an $h/2e$ vortex across a wire leads to a $2\pi$ phase-slip, which in turn switches the ground-state fermion-parity. Without creating additional excitations, a single vortex can only move across two wires at a time by exchanging a fermion between the wires, as shown in (a).}}
  \label{WTSC}
\end{figure}

The 2D weak TSC may be viewed as a stack of 1D TSCs (i.e. Kitaev wires) coupled by Josephson tunneling and requires the presence of translation symmetry, as illustrated in Fig.~\ref{WTSC} (a). Recall first that, a conventional $s$-wave superconductor possesses a $\mathbb{Z}_2$ topological order, provided that flux-binding vortices are treated as dynamical excitations. This requires including a fluctuating electromagnetic gauge field in the theory~\cite{Hansson2004SCisTO,Radzihovsky2017TOinSC}. Since the Bogoliubov quasiparticle and the $h/2e$ vortex have mutual $\pi$-braiding statistics \cite{Reznik89}, a connection with Kitaev's toric code can be established by identifying the former as the $f$-particle \footnote{This is originally called the $\varepsilon$-particle in \citep{Kitaev2006exactly}, but here we adopt another conventional label $f$ to emphasize its fermionic nature.} and the latter as the $m$-particle, such that their composite is the $e$-particle. In the case of a weak TSC, an additional crucial property arises, which originates from the fractional Josephson effect in 1D TSCs (see Fig.~\ref{WTSC} (b,c)). In particular, a $2\pi$ phase slip in a 1D TSC must lead to a change of the ground-state fermion-parity \cite{KitaevChain, Kwon2004, Cheng2015}. Since the 2D weak TSC can be considered as an array of 1D TSC wires, an $h/2e$ vortex is forbidden (by an energetic cost) to tunnel across a \textit{single} wire, but is allowed to tunnel across \textit{two} wires. The latter must occur by exchanging an $f$-fermion between the two neighboring wires, as depicted in Fig. \ref{WTSC}(a). Consequently, \textit{two} types of $h/2e$ vortices must be distinguished: vortices living on \emph{odd} links ($m_o$) and vortices living on \emph{even} links ($m_e$). Note that the braiding between $m_o$ and $m_e$ involves moving an $f$-particle around an $m$-particle, which implies a mutual $\pi$-braiding between the two. Hence $m_o/m_e$ can be respectively identified with the $e/m$-anyon in a toric code. Such a toric code topological order is symmetry-enriched in the sense that anyons $e$ and $m$ are related by a translation symmetry.

These general arguments show that a symmetry-enriched topological phase akin to the toric code can arise in a weak TSC coupled to a dynamical electromagnetic field. The purpose of this paper is to demonstrate that a different but related symmetry-enriched topological phase can be realized in an electrical insulator---which we refer to as the ``toric code insulator"---by condensing double-vortices which bind $h/e$-flux. To correctly describe the dynamical quantum excitations in the insulator, it is \textit{not} required to include the electromagnetic field. Instead, we will show that an emergent $\mathbb{Z}_2$ gauge field naturally arises in a coupled wire model for the interacting weak TSC. This results in a three-fluid wire model for the toric code insulator, with two fluids associated to the weak TSC and one fluid for the $\mathbb{Z}_2$ gauge field. The remnant of a single $h/2e$ vortex exists as a $\mathbb{Z}_2$ gauge flux.

This $\mathbb{Z}_2$ gauge structure is similar in spirit to the Senthil-Fisher theory for a $\mathbb{Z}_2$ fractionalized insulator \cite{SenthilFisher, SenthilFisher2001}, where the $\mathbb{Z}_2$ gauge flux (referred to as the ``vison") plays the role of a toric code anyon. It was pointed out that the $\mathbb{Z}_2$ fractionalized insulator proposed by Senthil and Fisher can also result from double-vortex condensation \cite{Balents1999}. Importantly, however, the toric code insulator considered here has \textit{two} types of visons: $\mathbf{m}_e$ and $\mathbf{m}_o$, which are remnants of the unpaired $h/2e$ vortices in the 2D weak TSC. As we will demonstrate, the two types of visons have a mutual $\pi$-braiding and are related by translation symmetry, thus realizing a symmetry-enriched toric code in an insulator. While our model consists of spinless electrons, there exists a neutral fermionic quasiparticle, analogous to the ``spinon", which arises from electron fractionalization and carries a $\mathbb{Z}_2$ gauge charge. After establishing a microscopic model, we shall elaborate on this connection to the Senthil-Fisher theory and comment on possible realizations in spinful electronic systems, with either spontaneously or explicitly broken time-reversal symmetry. 

\subsection{Outline of the paper}
The rest of the paper is organized as follows. In Sec. \ref{sec2} we present a microscopic model for the toric code insulator using the coupled wire construction. In Sec. \ref{sec3.1}, the anyonic spectrum and braiding statistics are analyzed. In particular, by constructing local operators that transport anyons, we discover a constrained motion for the visons that lead to symmetry-enrichment. An interesting consequence is a size-dependent ground state degeneracy on torus, which we elucidate in Sec. \ref{sec3.2}. In Sec. \ref{sec4}, we provide a survey of possible material platforms for realizing the toric code insulator, and discuss a band-theoretic perspective that connects to recent proposals of attaining fractionalization in semimetals. We conclude in Sec. \ref{conclusion}.

\section{\label{sec2} Wire model}

Our approach to constructing the toric code insulator relies on a coupled wire model for an array of spinless single-channel quantum wires. Since the 2D weak TSC can be viewed as a stack of 1D TSCs, the coupled wire model provides a natural description for the weak TSC, and as we will see, it also provides a description for the double-vortex condensation that leads to the toric code insulator. In this section, we begin by introducing a two-fluid model for a single wire. Then, by considering the competition between the inter-wire Josephson coupling and the intra-wire charge-density wave ordering, a $\mathbb{Z}_2$ gauge structure emerges. This leads us to a three-fluid wire model for a \textit{disordered} weak TSC, which is the symmetry-enriched toric code insulator.

\subsection{Two-fluid model for a single wire}\label{sec2.1}

Let us first review a bosonized theory for a 1D TSC, which is developed in Ref.~\onlinecite{CWC17}. A single 1D TSC can be described by a two-fluid model, where a Luttinger liquid of charge-$e$ fermions (i.e. the spinless electrons) coexists with a Luttinger liquid of charge-$2e$ bosons (i.e. the Cooper pairs). Using Abelian bosonization \cite{giamarchi2003, gogolin2004}, the right/left-moving ($R/L$) electron operator and the Cooper pair operator ($\Psi_{2e} $) are expressed as \footnote{Notice that Klein factors can be safely dropped here. In the subsequent discussion, the only place where Klein factors should arise is in the pairing interaction, i.e. $\psi^R_e \psi^L_e \Psi^\dagger_{2e}$ , while all other terms to be considered in the Hamiltonian are intrinsically bosonic. The Klein factor thus commutes with the Hamiltonian, and does not affect our central discussion.}
\begin{equation}\label{electronop}
\psi^{R/L}_e = e^{i(\frac{1}{2}\varphi_1\pm \theta_1)}, \quad \Psi_{2e} = e^{i\varphi_2},
\end{equation}
and their minimal density fluctuations are $e^{2i\theta_1}$ and $e^{i\theta_2}$ respectively. Here, the variables $\varphi_\alpha$ and $\theta_\alpha$ ($\alpha=1,2$) are canonically conjugate with the commutator $[ \partial_x \theta_\alpha (x), \varphi_\beta (x')] = 2\pi i \delta_{\alpha\beta} \delta(x-x')$. In our convention, the $x$-direction is along the wire (hence continuous), and later we will stack up an array of wires in the $y$-direction. It is convenient to define the charge- and neutral-sector variables as follows:
\begin{subequations}
\begin{align}
\varphi_\rho &=\varphi_2, &&\theta_\rho = \theta_1+\theta_2,\\
\varphi_\sigma &= \varphi_1-\varphi_2, &&\theta_\sigma  = \theta_1.
\end{align}
\end{subequations}
Note that the commutation relations are preserved and that the charge-sector variable $\varphi_\rho$ is the superconducting phase. The Hamiltonian of a single wire is chosen to be $\mathcal{H}_\text{wire}  = \mathcal{H}_\rho+\mathcal{H}_\sigma$, with
\begin{subequations}
\begin{align}
\mathcal{H}_\rho &= \frac{v_\rho}{4\pi} [g_\rho(\partial_x\varphi_\rho)^2+\frac{1}{g_\rho}(\partial_x \theta_\rho)^2],\\
\mathcal{H}_\sigma &=\frac{v_\sigma}{4\pi}[g_\sigma(\partial_x\varphi_\sigma)^2+\frac{1}{g_\sigma}(\partial_x \theta_\sigma)^2]+ u\cos \varphi_\sigma +v\cos 2\theta_\sigma.
\end{align}
\end{subequations}
Here, $g_{\rho,\sigma}$ are the Luttinger parameters of the charge and neutral Luttinger liquids. The term with coupling constant $u$ describes the pairing interaction between the two fluids, which turns two electrons into a Cooper pair and vice versa. The interaction with coupling constant $v$ describes the single electron back-scattering.

In the limit where $v \gg u$, the unpaired electrons are depleted by back-scattering and the wire becomes a trivial 1D superconductor which is gapless only to two-electron excitations. Instead, in the opposite limit $u \gg v$ the electrons are weakly-paired in the sense that charge-$e$ and charge-$2e$ fluid coexist, and the wire is a topological superconductor (with a fluctuating phase). As pointed out in Ref.~\onlinecite{CWC17}, the weakly paired phase is adiabatically connected to a single-channel Luttinger liquid with attractive interactions, and is gapless to both one-electron and two-electron excitations. The gapless charge-$e$ excitation corresponds to a composite operator given by
\begin{equation}
\psi_{\pm} = e^{i[\frac{1}{2}(\varphi_\rho+\varphi_\sigma) \pm \theta_\rho]}, \label{composite}
\end{equation}
which is a composite of adding a bare electron and tunneling a vortex across the wire (because $e^{\pm i\theta_\rho}$ introduces a $\pm 2\pi$ phase-slip in $\varphi_\rho$). The gaplessness of this composite one-electron excitation simply reflects the fractional Josephson effect in a 1D TSC: a $\pm 2\pi $ phase-slip switches the fermion-parity of the ground state.


As a next step, we couple an array of 1D TSCs by Josephson-tunneling between neighboring wires, resulting in a 2D weak TSC, and then consider phase-disordering the superconductor. We label the wires by $j$ and links between the wires by $\ell $, which is related to wire label as $\ell = j+1/2$. The Josephson coupling (coupling constant $J$) between the wires can then be expressed as
\begin{equation}
\mathcal{H}_{\text{SC}} = J \sum_\ell \cos (\Delta_y\varphi_\rho)_\ell, \label{josephson}
\end{equation}
where we have defined the discrete ``derivative'' $\Delta_y$ of wire variables as $\Delta_y\varphi_\rho \equiv \varphi_{\rho,j+1}-\varphi_{\rho,j} $. The $\Delta_y$ derivative is associated with link $\ell=j+1/2$ and we thus write $(\Delta_y\varphi_\rho)_\ell$. A vortex with winding number $n$, which binds $nh/2e$-flux, corresponds to a $2n\pi$-kink in $\Delta_y\varphi_\rho$ (and thus lives on the links). Hence, a pinned charge-density wave (CDW) on wire $j$ of the form
\begin{equation}
\mathcal{H}_{\text{CDW},j} = w \cos 2\theta_{\rho,j}
\end{equation}
competes with $\mathcal{H}_\text{SC}$ (since $\varphi_\rho$ and $\theta_\rho$ are conjugate variables) by tunneling \emph{double} vortices across wire $j$. In the large-$w$ limit, double vortices are condensed and superconductivity is destroyed by the rapid phase fluctuations. However, since double vortices correspond to $\pm4\pi$-kinks, there still remains a binary degree of freedom that distinguishes configurations with a uniform $\Delta_y \varphi_\rho$ from those with $2\pi$-kinks in $\Delta_y \varphi_\rho$. The latter are remnants of the uncondensed single vortices. To properly treat these as dynamical quantum excitations and to study the properties of the insulating ground state, we now introduce a $\mathbb{Z}_2$ gauge theory, which will be treated as the third fluid in the wire model.

\subsection{$\mathbb{Z}_2$ gauge theory}\label{sec2.2}

To see how a $\mathbb{Z}_2$ gauge sector emerges from the Luttinger  liquids, consider the compactification of the fields $\varphi_1$ and $\varphi_2$. In the above formulation, both fields are defined on a circle such that $\varphi_1 \equiv \varphi_1 + 4\pi$ and $\varphi_2 \equiv \varphi_2 + 2\pi$. As mentioned, in this case the minimal density operators are $e^{2i\theta_1}$ and $e^{i\theta_2}$, respectively. Given the compactification of $\varphi_2$, a $2\pi$ phase winding of $\varphi_2$ corresponds to a superconducting vortex. In terms of the charged and neutral variables, however, a $2\pi$-shift of $\varphi_2$ implies a \textit{simultaneous} $2\pi$-shift of $\varphi_\rho$ and $\varphi_\sigma$, from which one identifies $(\varphi_\rho, \varphi_\sigma) \equiv (\varphi_{\rho}+2\pi, \varphi_\sigma -2\pi)$. The latter is reflected, for instance, in the form of the composite operator in Eq.~\eqref{composite}. In this sense the charged and neutral variables are coupled, and \textit{not} independent. 

\begin{figure}[b]
   \includegraphics[width=\columnwidth]{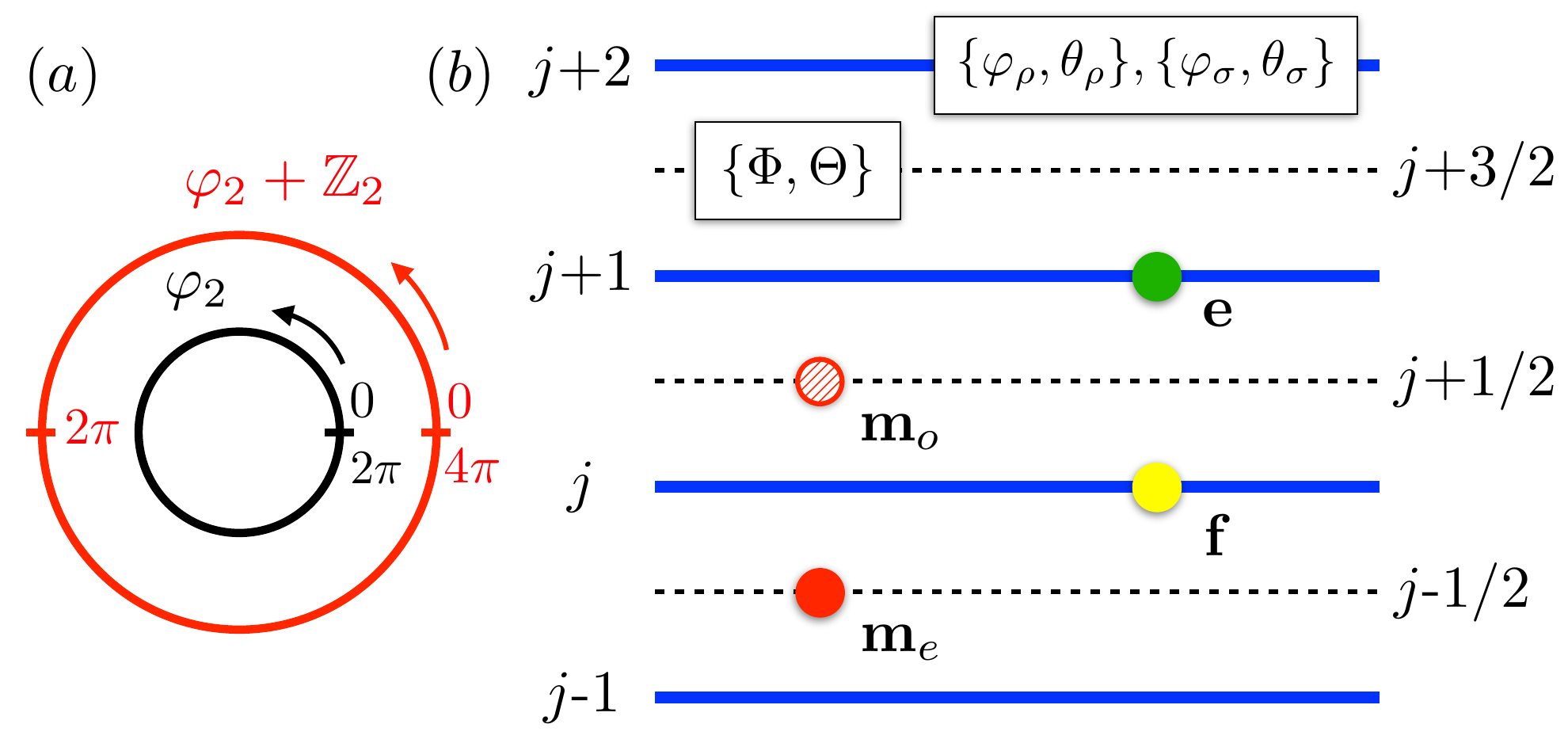}\centering
  \caption{\small{(a) Two formulations for the Luttinger liquid of Cooper pairs. In one, $\varphi_2$ has $2\pi$ compactification. Alternatively, $\varphi_2$ can have $4\pi$ compactification and couple to a $\mathbb{Z}_2$ gauge symmetry: $\varphi_2 \mapsto \varphi_2 +2\pi$. (b) Three-fluid wire model of the toric code insulator: two fluids for the charge and neutral sectors of a weak TSC, while the third fluid describes the $\mathbb{Z}_2$ gauge sector emerging from double-vortex condensation. Anyons from each sector are labeled.}} 
  \label{wiremodel}
\end{figure}

A $\mathbb{Z}_2$ gauge theory is introduced by compactifying $\varphi_2$ on a \emph{larger} circle, see Fig. \ref{wiremodel}(a), such that $\varphi_2 \equiv \varphi_2 + 4\pi$, and coupling $\varphi_2$ to a $\mathbb{Z}_2$ gauge field which mods out the shift symmetry $\varphi_2 \mapsto \varphi_2 + 2\pi$. In this formulation, the minimal density operator becomes $e^{2i\theta_2}$, while $e^{i\theta_2}$ is replaced by a twist operator, which tunnels single vortices across a wire. The vortices themselves correspond to a non-trivial $\mathbb{Z}_2$ magnetic flux, which is defined on the links and can hop across a wire by applying the electric field operator $e^{iE_x/2}$. The latter is the new twist operator, to be considered in more detail below. This formulation, in which the single vortices are represented as the magnetic flux of a $\mathbb{Z}_2$ gauge field, is equivalent to the original formulation. It has the benefit that the charge and neutral variables are treated as independent, but are coupled via the gauge field. In this sense, our approach is similar in spirit to previous approaches, such as slave-particle representations, which introduce a gauge symmetry to liberate new degrees of freedom \cite{SenthilFisher}.

To further develop the \textit{gauged} two-fluid model, it is instructive to consider the emergence of the $\mathbb{Z}_2$ gauge field at the level of the partition function. In particular, consider the Josephson tunneling term \eqref{josephson} in the partition function, which may be rewritten using Villain's method as \cite{kleinert1989gauge}
\begin{equation}\label{Z2Villain}
\begin{split}
e^{-J \cos \Delta_y\varphi_\rho} &\sim \sum_{n\in \mathbb{Z}}e^{\frac{J}{2}(\Delta_y\varphi_\rho-2n\pi)^2}\\
& \sim \sum_{A_y=0,\pi} e^{-4J\cos(\Delta_y\varphi_\rho/2-A_y)}.
\end{split}
\end{equation}
In the first line the cosine is approximated using Villain's prescription, which is reversed in the second line after rearranging the sum over integers into separate sums over even and odd integers. By re-expressing the partition function in this way, we have introduced a $\mathbb{Z}_2$ gauge field $A_y \in \{0,\pi\}$, which is defined on the links, as is $\Delta_y\varphi_\rho$, and enters as a minimal coupling to the superconducting phase. As is evident from \eqref{Z2Villain}, a $2\pi$ phase-slip in $\Delta_y\varphi_\rho$ can be compensated by a $\pi$-kink in $A_y$, which reflects the aforementioned identification of magnetic gauge flux with the superconducting vortex. It is worth noting that Eq.~\eqref{Z2Villain} is similar in spirit to the $\mathbb{Z}_2$ gauge theory developed by Senthil and Fisher~\cite{SenthilFisher}, which they introduced by splitting the Cooper pair in half (i.e., $\varphi \rightarrow \varphi/2$), at the expense of a gauge degree of freedom.  

Gauging the $\mathbb{Z}_2$ symmetry on individual wires leads to the minimal coupling of the gauge field components $A_x$ and $A_t$ to the phase variables $\varphi_2$: $\partial_{t,x}\varphi_2 - 2A_{t,x}$. Analogous to Maxwell's $U(1)$ gauge theory, one can define a magnetic field $B = \partial_x A_y- \Delta_y A_x$, and conjugate electric fields $(E_x,E_y)$ that satisfy the canonical commutation relations:
\begin{subequations}
\begin{align}
[E_{x,j}(x),A_{x,j'}(x')] &=i2\pi \delta_{jj'}\delta(x-x'),\\ 
[E_{y,\ell}(x),A_{y,\ell'}(x')] &=i2\pi \delta_{\ell\ell'}\delta(x-x').
\end{align}
\end{subequations}
It is clear from the above that $e^{iE_x/2}$ is a local operator that tunnels a $\pi$-magnetic flux across a single wire. This operator will play an important role when we discuss anyon statistics and symmetry-enrichment in the next section.

The above 2+1D gauge theory can be formulated as an array of Luttinger liquids defined on links, each characterized by a pair of canonically conjugate bosonic fields $\Theta_\ell(x)$ and $\Phi_\ell(x)$, with $\Theta_\ell$ having a $2\pi$-compactification and $e^{i\Phi_\ell}$ being an allowed operator. To that end, we identify $A_{y,\ell} = \Theta_\ell$, and consider a sine-Gordon Hamiltonian with a large $\cos2\Theta_\ell$ to impose the discreteness of the gauge field. We then adopt the $A_{x,j}=0$ gauge, and express:
\begin{subequations}\label{EM}
\begin{align}
B_\ell &=\partial_x \Theta_\ell , \;\; E_{y,\ell} = \partial_x \Phi_\ell, \label{EMfields}\\
E_{x,j} &= 2(\theta_{\rho}-\theta_{\sigma})_j-(\Delta_y \Phi)_j. \label{Gausslaw}
\end{align}
\end{subequations}
As such, a $\pi$-magnetic flux on link-$\ell$ corresponds to a $\pi$-kink in $\Theta_\ell$. The discrete derivative $\Delta_y$ for link variables is defined as $(\Delta_y \Phi)_j \equiv \Phi_{j+1/2}-\Phi_{j-1/2}=\Phi_{\ell}-\Phi_{\ell-1}$. The expression for $E_{x}$ (defined modulo $4\pi$) is the Gauss's law constraint for the physical Hilbert space, which can be obtained from integrating out $A_t$ in the theory. The Gauss's law constraint is particularly important here, as the twist operator $e^{iE_x/2}$ needs to be recast into a form that is compatible with the $A_{x,j}=0$ gauge, under which the Luttinger liquid representation of the gauge theory is constructed. A derivation for the Gauss's law constraint is presented in Appendix \ref{AppA}.\\

\setlength{\tabcolsep}{1em}
\begin{table*}[t]
\centering
\begin{tabular}{ c | c | c | c | c | c }
Anyon & Symbol & Operator & Hopping operator & Self-statistics & Mutual $\pi$-braiding with \\
 \hline
Neutral fermion & $\mathbf{f}$ & $e^{i(\frac{1}{2}\varphi_{\sigma}+\theta_{\sigma})_j}$ & $e^{i(\frac{1}{2}\Delta_y\varphi_\sigma+\Delta_y\theta_\sigma +\Theta)_\ell}$ & Fermion & $\mathbf{m}_e$\;,\;$\mathbf{m}_o$\\
Chargon & $\mathbf{e}$ &  $e^{\frac{i}{2}\varphi_{\rho,j}}$ & $e^{i(\frac{1}{2}\Delta_y \varphi_\rho-\Theta)_\ell}$ & Boson & $\mathbf{m}_e$\;,\; $\mathbf{m}_o$\\
Vison (even-link $\ell$) & $\mathbf{m}_e$ &  $e^{\frac{i}{2}\Phi_{\ell}}$ & see Eq. (\ref{Tm}) & Boson & $\mathbf{m}_o$\;, $\mathbf{f}$\;,\; $\mathbf{e}$\\
Vison (odd-link $\ell$) & $\mathbf{m}_o$ &  $e^{\frac{i}{2}\Phi_{\ell}}$ &  see Eq. (\ref{Tm}) & Boson & $\mathbf{m}_e$\;, $\mathbf{f}$\;,\; $\mathbf{e}$\\
\end{tabular}
\caption{Summary of the spectrum and anyon statistics in the toric code insulator, which exhibits charge-statistics separation and symmetry-enriched topological order. Neutral fermions and chargons live on wires (labeled by $j$), while visons live on links (labeled by $\ell=j+1/2$), with even/odd-link corresponding to even/odd integer $j$. The local hopping operator moves an anyon across one link, along the direction perpendicular to the wires. In particular, without creating additional excitations, a vison has to hop across two wires at a time, thus differentiating between $\mathbf{m}_e$ and $\mathbf{m}_o$.}
\label{spectrum}
\end{table*}

\subsection{Three-fluid model for coupled wires}\label{sec2.3}

The stage is now set for constructing a microscopic model for the symmetry-enriched toric code insulator. This is a three-fluid model describing a disordered weak TSC, as sketched in Fig. \ref{wiremodel}(b), with the Hamiltonian: $\mathcal{H} = \mathcal{H}_{\sigma}+\mathcal{H}_{\rho}+\mathcal{H}_{\mathbb{Z}_2}$. Here,
\begin{subequations}\label{3fluidH}
\begin{align}
\mathcal{H}_{\sigma} &= \sum_j \frac{v_\sigma}{4\pi} \big[ g_\sigma (\partial_x \varphi_{\sigma,j})^2+\frac{1}{g_\sigma}(\partial_x \theta_{\sigma,j})^2  \big]+ u\cos\varphi_{\sigma,j}, \\
\mathcal{H}_{\rho} &= \sum_j \frac{v_\rho}{4\pi} \big[ g_\rho (\partial_x \varphi_{\rho,j})^2+\frac{1}{g_\rho}(\partial_x \theta_{\rho,j})^2  \big]+ w\cos 2\theta_{\rho,j}, \\
\begin{split}
\mathcal{H}_{\mathbb{Z}_2} &=  \sum_\ell \frac{V}{4\pi}\big[ g (\partial_x \Phi_{\ell})^2+\frac{1}{g}(\partial_x \Theta_{\ell})^2  \big] + h \cos 2\Theta_{\ell}.
\end{split}
\end{align}
\end{subequations}
Three important ingredients are in place: the $u$-term makes the wires into 1D TSCs, the $w$-term introduces a CDW that disorders the 2D superconductivity which would otherwise be obtained from Josephson coupling, and the $h$-term implements the remaining gapped $\mathbb{Z}_2$ degree of freedom. When all the interactions (i.e. $u,w,h$) are large, the system is a gapped insulator. Despite the apparent decoupling of the three fluids in $\mathcal{H}$, they are indeed coupled given the set of allowed local operators, and as we show next, all these conspire to give a non-trivial state exhibiting a charge-statistics separation and a symmetry-enriched $\mathbb{Z}_2$ topological order.\\

\section{\label{sec3} Translation symmetry-enriched topological order}

The central result of this work lies in the physical features of the insulating state constructed in the previous section and concerns the quasiparticle excitations, i.e., $2\pi$-kinks in the argument of the interacting cosine terms. For reasons that will become clear, we refer to the excitations from the $u$-, $w$- and $h$-terms as ``neutral fermion" ($\mathbf{f}$), ``chargon" ($\mathbf{e}$) and ``vison" ($\mathbf{m}$) respectively. They are also labeled in Fig. \ref{wiremodel}(b). As will be explained below, they are actually anyons with non-trivial braiding statistics. Importantly, there are in fact two types of visons: $\mathbf{m}_o$ and $\mathbf{m}_e$, which are mutual-semions that altogether form a toric code topological order enriched by translation symmetry. 

\subsection{\label{sec3.1} Anyon spectrum and statistics}

Let us first begin with the more evident features. Given the allowed local operators $e^{2i\theta_{\sigma,j}}$ and $e^{i\Phi_\ell}$, which create $4\pi$-kinks in the argument of the $u$- and $h$-cosine terms respectively, we conclude that both the neutral fermion and the vison obey a $\mathbb{Z}_2$ fusion rule:
\begin{equation}
\mathbf{f}^2 = \mathbf{m}^2= \mathds{1}.
\end{equation}
From the charge-density fluctuation on wire-$j$: $\rho_j(x) = \partial_x \theta_{\rho,j} (x) /\pi$, we conclude that a chargon (which corresponds to a $\pi$-kink in $\theta_{\rho,j}$) carries an electric charge-$e$ of a single electron, hence labeled as the $\mathbf{e}$-particle.

Next we study the anyonic motion. The anyon operators are summarized in Table. \ref{spectrum}. From the canonical commutation relations $[\varphi (x) ,\theta(x')]$ and $[\Phi(x), \Theta(x')]$, they can be easily checked to create $2\pi$-kinks in the corresponding interaction terms. Since anyons are non-local objects, a single one of them cannot be created/annihilated by local operators; however, they can be \textit{pair-created} from the ground state by acting local operators, and this action can be equivalently interpreted as \textit{moving} the anyon (i.e. let it be annihilated in one place and created in somewhere else). These operators are referred to as the hopping operators, and their explicit form dictates the braiding statistics of the anyons. The hopping operators for motion along $\hat{x}$ are easily constructed. For instance, a chargon can move along the wire by
\begin{equation}
\mathbf{e}_j(x_1) \;\mathbf{e}^\dagger_j(x_2) = \exp[\;\frac{i}{2}\int^{x_1}_{x_2}dx\;\partial_x \varphi_{\rho,j}\;],
\end{equation}
which is indeed a local operator as $\partial_x \varphi_\rho$ represents the current operator in the Abelian bosonization. Similarly, one can argue that $\mathbf{f}_j(x_1) \mathbf{f}^{\dagger}_j(x_2)$ is a local operator. Visons, which are created by $e^{\frac{i}{2}\Phi_\ell}$, can move along $\hat{x}$ by applying the $y$-component of the electric field operator, namely
\begin{equation}
\begin{split}
\mathbf{m}_\ell(x_1)\;\mathbf{m}^\dagger_\ell(x_2) &= e^{\frac{i}{2}[\Phi_\ell(x_1)-\Phi_\ell(x_2)]} \\
&= \exp[\;\frac{i}{2}\int_{x_2}^{x_1}E_{y,\ell}\;dx\;].
\end{split}
\end{equation}

\begin{figure}[t]
   \includegraphics[width=1\columnwidth, height=0.8\columnwidth ]{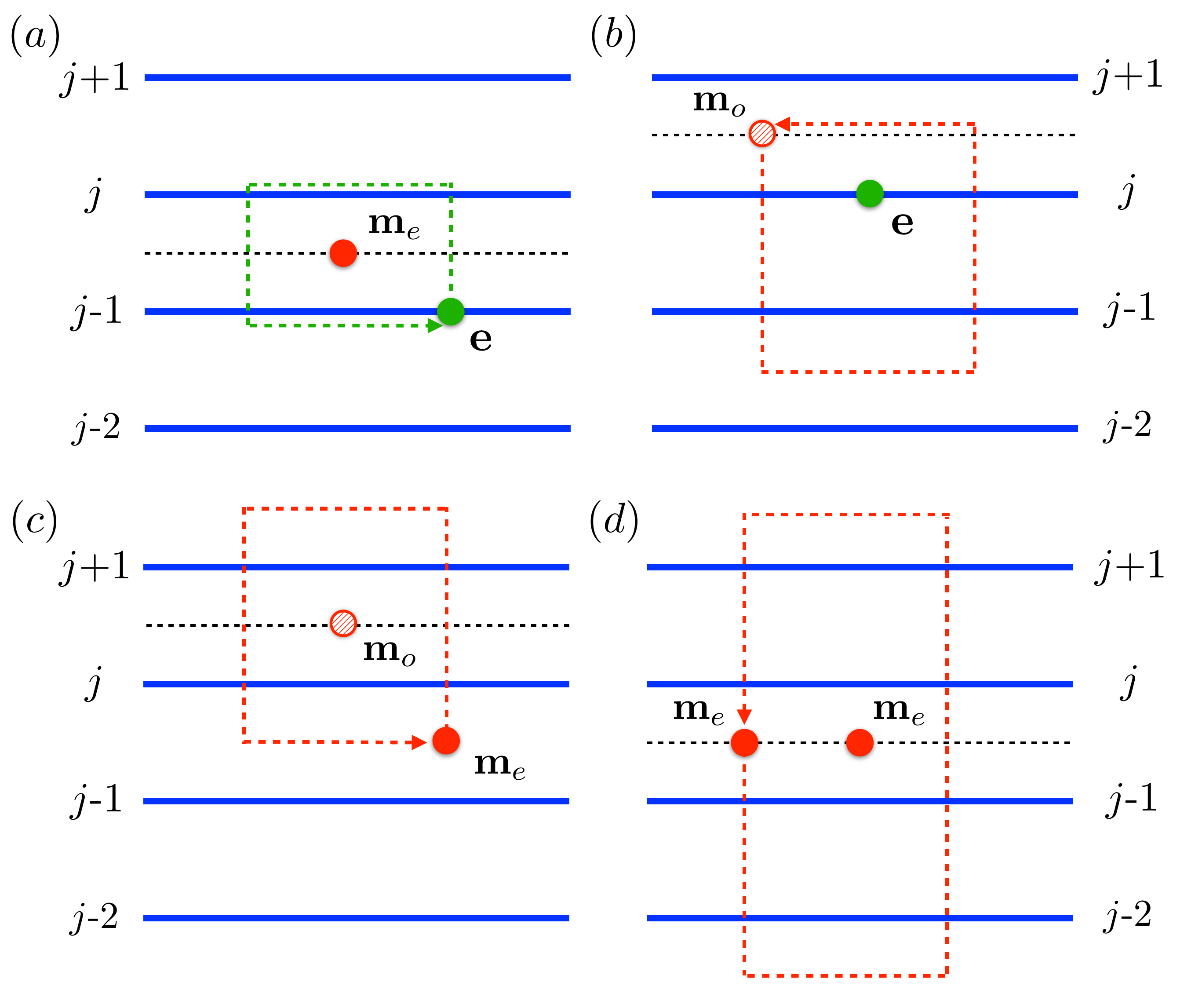}\centering
  \caption{\small{Representative braiding processes in the toric code insulator. The braiding phase can be computed by multiplying a string of operators that transport the anyon around a closed loop, depicted by a dashed line in each panel. (a,b): Braiding between $\mathbf{e}$ and $\mathbf{m}$ leads to a $\pi$-phase. The same happens for the braiding between $\mathbf{f}$ and $\mathbf{m}$, as $\mathbf{f}$ is topologically equivalent to $\mathbf{e}=\psi_e \times \mathbf{f}$. (c): Braiding between $\mathbf{m}_o$ and $\mathbf{m}_e$ leads to a $\pi$-phase. (d): The self-statistics of $\mathbf{m}_\text{even/odd}$ is trivial. All these can be deduced from the expression of operator $T_\ell(\mathbf{m})$, which transports an $\mathbf{m}$-particle across two wires.}}
  \label{braidings}
\end{figure}

The more interesting hopping operators are for the motion perpendicular to the wires, which explicitly encode information about the braiding and exchange statistics. From the partition function in Eq. (\ref{Z2Villain}), along with the identification of $A_{y,\ell} = \Theta_\ell$, we can identify the local operator that transports a chargon across link-$\ell$:
\begin{equation}\label{Te}
T_\ell(\mathbf{e})  = e^{i(\frac{1}{2}\Delta_y \varphi_\rho-\Theta)_\ell}.
\end{equation}
By composing this with the electron hopping operator $\exp i\Delta_y(\frac{1}{2}\varphi_\rho+\frac{1}{2}\varphi_\sigma+\theta_\sigma)$, we obtain the local operator that transports an $\mathbf{f}$-particle across link-$\ell$:
\begin{equation}\label{Tf}
T_\ell(\mathbf{f})  = e^{i(\frac{1}{2}\Delta_y \varphi_\sigma+\Delta_y\theta_\sigma+\Theta)_\ell}.
\end{equation}
We first notice that the above combination $e^{i(\frac{1}{2}\varphi_\sigma +\theta_\sigma)}$ for the $\mathbf{f}$-particle suggests that it, as a $2\pi$-kink in $\varphi_\sigma$, is a self-fermion, since $e^{\frac{i}{2}\varphi_\sigma}$ is acting as a Jordan-Wigner string. In contrast, the charge-$e$ $\mathbf{e}$-particle is a self-boson, since $\theta_\rho$ does not show up in Eq. (\ref{Te}). The above considerations suggest that an electron can \textit{fractionalize} into a bosonic chargon (which carries charge-$e$ of the electron) and a neutral fermion (which carries the fermionic self-statistics of the electron) in the toric code insulator. This resembles the fractionalization noticed by Senthil and Fisher in the $\mathbb{Z}_2$ gauge theory of cuprate superconductors \cite{SenthilFisher}, hence a similar nomenclature has been adopted here. More explicitly, the bare electron operator in Eq. (\ref{electronop}) can be written as
\begin{equation}\label{separation}
\psi_e = e^{\frac{i}{2}\varphi_\rho} \times e^{i(\frac{1}{2}\varphi_\sigma+\theta_\sigma)} =\mathbf{e} \times \mathbf{f}.
\end{equation}
There is a local $\mathbb{Z}_2$ symmetry that transforms $(\varphi_\rho,\varphi_\sigma) \mapsto (\varphi_\rho+2\pi,\varphi_\sigma+2\pi)$, or equivalently
\begin{equation}
\mathbb{Z}_2:\quad (\mathbf{e},\mathbf{f}) \mapsto (-\mathbf{e}, -\mathbf{f}), 
\end{equation}
which leaves the electron operator invariant. This is the origin of the $\mathbb{Z}_2$ gauge field that we introduce in Sec. \ref{sec2.2}, which couples to both the chargon and the neutral fermion. Its gauge flux is known as the ``vison". From Eqs. (\ref{Te}) and (\ref{Tf}), one should notice a phase factor $e^{\pm i\Theta_\ell}$ being picked up by transporting $\mathbf{e/f}$. Since a vison on link-$\ell$ corresponds to a $\pi$-kink in $\Theta_\ell$, this implies a $\pi$-braiding between $\mathbf{e}/\mathbf{f}$ and $\mathbf{m}$, as illustrated in Fig. \ref{braidings}(a). 

Despite the above similarities with the Senthil-Fisher model, the toric code insulator just constructed has an additional feature of symmetry-enrichment. To see that, let us analyze how visons move along $\hat{y}$, and derive a mobility constraint. A local operator that hops a $\pi$-kink in $\Theta_\ell$ can be obtained from the twist operator, $e^{iE_{x,j}/2}$, introduced in Sec. \ref{sec2.2}. From the Gauss's law constraint in Eq. (\ref{Gausslaw}), we have
\begin{equation}\label{twist}
e^{\frac{i}{2}E_{x,j}} = e^{i(\theta_\rho-\theta_\sigma-\frac{1}{2}\Delta_y\Phi)_j},
\end{equation}
hence the right-hand side provides a physical operator in the our chosen gauge: $A_{x}=0$. However, this operator not only hops a vison (by $e^{-\frac{i}{2}\Delta_y\Phi}$), but also \textit{creates} an $\mathbf{f}$-particle, which is a $2\pi$-kink in $\varphi_\sigma$ as created by $e^{-i\theta_{\sigma}}$. This is the reincarnation of the fractional Josephson effect: an $\mathbf{m}$-particle simply cannot hop across a single wire without changing the fermion-parity of the wire. Nevertheless, by applying the above twist operator \textit{twice}, a vison can hop across two neighboring wires ($j$ and $j+1$), or equivalently across a link ($\ell=j+1/2$), \textit{without} creating excitations as long as the hopping operator in Eq. (\ref{Tf}) is \textit{also} applied to annihilate additional $\mathbf{f}$-particles. Namely, we consider the local operator
\begin{equation}\label{Tm}
\begin{split}
T_{\ell} (\mathbf{m}) &= e^{\frac{i}{2}(E_{x,j}+E_{x,j+1})} \times  T_{\ell}(\mathbf{f}) \times e^{2i\theta_{\sigma,j}}\\
&= e^{i(\frac{1}{2}\Delta_y\varphi_\sigma +\Sigma_y\theta_\rho+\Theta)_\ell} \times e^{\frac{i}{2}(\Phi_{\ell-1}-\Phi_{\ell+1})}, 
\end{split}
\end{equation} 
where $(\Sigma_y \theta_{\rho})_\ell \equiv \theta_{\rho, j+1}+ \theta_{\rho,j}$. In the last line, the first part is a phase factor, and the second part has the action of tunneling a vison across two wires. We have made use of $\mathbf{f}^2= \mathds{1}$ by attaching a factor of $e^{2i\theta_{\sigma,j}}$, so as to completely eliminate $\theta_\sigma$ in the hopping operator, and hence no additional excitation is created during the tunneling process.

The upshots of the above analysis are three-fold: (1) there are two types of visons, one lives on the even-links ($\mathbf{m}_e$) and one lives on the odd-links ($\mathbf{m}_o$). While they are related by a translation symmetry, there is no local operator that turns one into the other without creating additional excitations. This leads to a pattern of \textit{weak symmetry breaking} akin to the one in Wen's plaquette model and Kitaev's honeycomb model \cite{Wenplaquette2003, Kitaev2006exactly, Rao2020}; (2) By the virtue of Eq. (\ref{twist}), a local operator can create a composite of three anyons: $\mathbf{m}_e$, $\mathbf{m}_o$ and $\mathbf{f}$. This implies the following fusion rule:
\begin{equation}\label{mmf}
\mathbf{m}_e \times \mathbf{m}_o = \mathbf{f}\;;
\end{equation}
Together with Eq. (\ref{separation}), this implies that a physical electron can fractionalize into one chargon $\mathbf{e}$ and two symmetry-related visons $\mathbf{m}_e$ and $\mathbf{m}_o$. (3) The braiding statistics related to the $\mathbf{m}$-particle is encoded in the phase factors in Eq. (\ref{Tm}), and schematically summarized in Fig. \ref{braidings}: the phase factor $e^{\frac{i}{2}(\Delta_y\varphi_\sigma)_\ell}$ implies a $\pi$-braiding between $\mathbf{m}$ and $\mathbf{f}$, $e^{i(\Sigma_y \theta_\rho)_\ell}$ implies a $\pi$-braiding between $\mathbf{m}$ and $\mathbf{e}$, and last but not least, $e^{i\Theta_\ell}$ implies a $\pi$-braiding between $\mathbf{m}_e$ and $\mathbf{m}_o$. Combining the above considerations, we have obtained a \textit{translation symmetry-enriched} toric code order, which is formed by $\{\mathbf{m}_e, \mathbf{m}_o, \mathbf{f}\}$. 

One signature of this symmetry-enrichment is revealed on the side edge, which is then described by a critical Ising-Majorana chain due to the correspondence between bulk anyonic symmetry and edge duality \cite{Lichtman2020bulkedge}. Alternatively, this can be seen in the 2D weak TSC, whose side edge is a chain of Majorana zero modes coupled in an \textit{undimerized} pattern, which leads to counter-propagating gapless Majorana fermions. Disordering the 2D WTSC by proliferating double-vortices in the bulk should not destroy the Majorana edge modes. The gapless neutral excitations might then be probed by thermal transport measurements \cite{Vogl2012}.

In summary, the toric code insulator constructed from disordering a 2D weak TSC is a tensor product between the symmetry-enriched toric code $\{\mathbf{m}_e, \mathbf{m}_o, \mathbf{f}\}$ and the physical electron $\psi_e$. The two types of visons, $\mathbf{m}_e$ and $\mathbf{m}_o$, can be respectively associated to the $e$-anyon and the $m$-anyon in Kitaev's toric code, and the $e \leftrightarrow m$ anyon permutation is realized here by a discrete translation that exchanges $\mathbf{m}_e \leftrightarrow \mathbf{m}_o$. Our model is different from Kitaev's toric code due to the presence of itinerant electron $\psi_e$, which additionally features charge-statistics separation: $\psi_e  = \mathbf{e} \times \mathbf{f}$. The $\mathbf{f}$-particle is a neutral fermion that can further fractionalize into $\mathbf{m}_e \times \mathbf{m}_o$, while the $\mathbf{e}$-particle is a charge-$e$ boson with a $\pi$-braiding with respect to both types of $\mathbf{m}$-visons.

\subsection{\label{sec3.2} Ground state degeneracy on torus: \\ an even-odd effect}
Related to the fusion and braiding properties of anyons, another prominent feature of a topological order is the ground state degeneracy (GSD) on a high genus Riemann surface. Below, we focus on the GSD on torus ($T^2$), by imposing periodic boundary conditions on both directions of the 2D system. For Kitaev's toric code, the GSD on torus is well-known to be 4 \cite{Kitaev2003TQC}. However, with a non-trivial interplay between translation symmetry and the topological order, there can be a size-dependent GSD as in the case of Wen's plaquette model \cite{Wenplaquette2003, Kou2008, Kou2009}. 

For the toric code insulator just constructed, the GSD depends on the parity of the number of wires, $L$, as follows:
\begin{equation}\label{GSD}
\text{GSD on } T^2 =
 \begin{cases} 
 4, \; \text{for even }L ; \\
 2, \; \text{for odd }L.
 \end{cases}
\end{equation}
The above result can be understood using a Wilson-loop argument. Let us consider the action of creating a pair of anyons from a ground state, then bringing one of the anyons all the way around a non-trivial cycle $\mathcal{C}_i$ ($i=x,y$) of $T^2$ and back to re-annihilate with its partner, finally returning the system back to a ground state. We denote the corresponding operator for the $\mathbf{a}$-anyon as $\mathcal{W}^{\mathbf{a}}_i$. Since the fundamental anyons are $\mathbf{m}_e$ and $\mathbf{m}_o$, while other excitations can be treated as composites of these (or together with the trivial electron), we shall focus just on the algebra generated by the operators associated to these two anyons. In the case of an even number of wires, from the mutual braiding statistics discussed above, we have:
\begin{subequations}
\begin{align}
\mathcal{W}^{\mathbf{m}_e}_x \; \mathcal{W}^{\mathbf{m}_o}_y = - \mathcal{W}^{\mathbf{m}_o}_y \; \mathcal{W}^{\mathbf{m}_e}_x, \\
\mathcal{W}^{\mathbf{m}_o}_x \; \mathcal{W}^{\mathbf{m}_e}_y = - \mathcal{W}^{\mathbf{m}_e}_y \; \mathcal{W}^{\mathbf{m}_o}_x, 
\end{align}
\end{subequations}
while other combinations of operators commute. The above algebra demands at the minimal a \textit{four}-dimensional ground-state Hilbert space: $\{ \ket{n,m} \mid n,m \in \mathbb{Z}_2 \}$, where the Wilson loop operators can be shown to act as: $\mathcal{W}^{\mathbf{m}_e}_x \ket{n,m} = \ket{n+1,m}$, $\mathcal{W}^{\mathbf{m}_o}_x \ket{n,m} = \ket{n,m+1}$, $\mathcal{W}^{\mathbf{m}_e}_y \ket{n,m} = (-1)^m \ket{n,m}$ and $\mathcal{W}^{\mathbf{m}_o}_y \ket{n,m} = (-1)^n \ket{n,m} $.

\begin{figure}[t]
   \includegraphics[width=0.6\columnwidth]{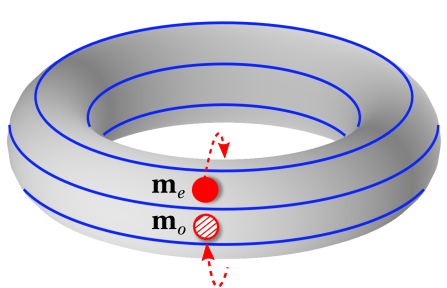}\centering
  \caption{\small{Toric code insulator with an odd number of wires. Due to the periodic boundary condition, an even-link vison ($\mathbf{m}_e$) turns into an odd-link vison ($\mathbf{m}_o$) after going around $\mathcal{C}_y$ once.}}
  \label{torusGSD}
\end{figure}

In case of an odd number of wires, the situation is quite different. Notice that by going around $\mathcal{C}_y$ \textit{once}, $\mathbf{m}_e$ and $\mathbf{m}_o$ are exchanged, as illustrated in Fig. \ref{torusGSD}. In other words,  $\mathcal{W}^{\mathbf{m}_e}_y$ and $ \mathcal{W}^{\mathbf{m}_o}_y$ are ill-defined, as the $\mathbf{m}$-particle cannot annihilate with its partner by just going around $\mathcal{C}_y$ once. Instead, we are forced to consider a new operator $\widetilde{\mathcal{W}}^\mathbf{m}_y$, which pair-creates two $\mathbf{m}_\text{e/o}$-particles and transports one of them around $\mathcal{C}_y$ \textit{twice}, and finally re-annihilates them. The Wilson operators that act on the ground-state subspace then obey the following relations (again derived from the braiding statistics):
\begin{subequations}
\begin{align}
\mathcal{W}^{\mathbf{m}_e}_x \; \widetilde{\mathcal{W}}^\mathbf{m}_y = -  \widetilde{\mathcal{W}}^\mathbf{m}_y\;\mathcal{W}^{\mathbf{m}_e}_x, \\
\mathcal{W}^{\mathbf{m}_o}_x \; \widetilde{\mathcal{W}}^\mathbf{m}_y = -  \widetilde{\mathcal{W}}^\mathbf{m}_y\;\mathcal{W}^{\mathbf{m}_o}_x,
\end{align}
\end{subequations}
while other combinations of operators commute. This time the minimal ground-state subspace is \textit{two}-dimensional: $\{ \ket{n} \mid n \in \mathbb{Z}_2 \}$, where the operators act as: $\mathcal{W}^{\mathbf{m}_e}_x \ket{n} = \mathcal{W}^{\mathbf{m}_o}_x \ket{n} = \ket{n+1}$ and $\widetilde{\mathcal{W}}^\mathbf{m}_y \ket{n} = (-1)^n \ket{n}$. In fact, one can show that $\mathcal{W}^{\mathbf{m}_e}_x$ and $\mathcal{W}^{\mathbf{m}_o}_x$ are related by a large gauge transformation in this case, so they are effectively the same operator which we shall denote as $\widetilde{\mathcal{W}}^{\mathbf{m}}_x$. The only non-trivial algebraic relation for the Wilson operators is that $\widetilde{\mathcal{W}}^{\mathbf{m}}_x$ \textit{anti-commutes} with $\widetilde{\mathcal{W}}^\mathbf{m}_y$, demanding a ground state subspace with minimal dimension 2. 

The Wilson-loop argument thus provides support to Eq. (\ref{GSD}), as a consequence of the translation symmetry-enrichment. In Appendix \ref{AppB}, we present a counting argument using the three-fluid wire model to explicitly demonstrate this even-odd effect, confirming that the minimal GSD is indeed realized in each case. In future numerical studies, this feature could be useful for identifying the symmetry-enriched toric code insulator.

\section{\label{sec4} Possible realizations}

We now discuss possible routes towards the experimental realization of a symmetry-enriched toric code insulator. Since the 2D weak TSC has provided the conceptual starting point for our analysis, we focus on three promising experimental platforms for realizing the weak TSC in two dimensions. A separate question concerns the precise mechanism by which vortex-condensation might occur, thus phase-disordering the superconductor and potentially giving rise to the toric code insulator. This question is intimately related to the nature of superconductivity, i.e., whether it is intrinsic or proximity-induced. We leave this question for future study, and instead focus more broadly on realizations of weak TSCs. 

In addition, we present a tight-binding model realization for the weak TSC on a square lattice. This serves as a toy model for certain available experimental platforms, and furthermore exposes an intriguing connection with strongly-interacting gapped semimetals, both in two and three dimensions.

\subsection{Relevant experimental platforms}

As a first example of a possible experimental platform, consider the side surface of a weak topological insulator (TI). The weak TI can be viewed as a stack of two-dimensional quantum spin Hall insulators protected by time-reversal symmetry (TRS) and translation symmetry in the stacking direction. This implies that a side surface, whose normal is perpendicular to the stacking direction, realizes a stack of counter-propagating helical (i.e., spin-momentum locked) edge modes, with one pair of counter-propagating modes per layer. Viewing the helical edge modes as quantum wires establishes a connection with the wire model introduced in the previous section, albeit with the important difference that the helical quantum wires can only exist as boundary modes of a topologically nontrivial bulk system protected by TRS. Breaking translation symmetry allows for the hybridization of two helical wires and leads to a \text{trivial} gapped surface. Opening up a pairing gap in a single helical wire gives rise to a topological superconducting phase similar to Kitaev's model for a spinless superconductor in one dimension. Notably, evidence for such a 1D phase has been recently reported in a WTe$_2$-NbSe$_2$ proximity-coupled heterostructure \cite{Lupke:2020p526}. This similarity with the Kitaev wire suggests that the physics of the 2D weak TSC can be emulated on the surface of a 3D weak TI. A promising experimental candidate is the recently reported weak TI bismuth iodine ($\beta$-Bi$_4$I$_4$), which has a quasi-1D structure and side surface states with weak dispersion in one of the two momentum directions \cite{Noguchi:2019p518}. As a result, our wire model may provide a useful starting point for describing such surfaces. 

The fate of the weak TI surface in the presence of strong interactions was explored in Ref.~\onlinecite{Mross:2016p036803}, which showed that a symmetric gapped surface is necessarily topologically ordered. The minimal Abelian order studied in  Ref.~\onlinecite{Mross:2016p036803} is of the $\mathbb{Z}_4$ type, whose fundamental anyons $a$ and $d$ are self-bosons and have mutual $\pi/2$-braiding statistics. In particular, in this $\mathbb{Z}_4$ topological order the $d$-particle transforms non-trivially under translation, very much like the $\textbf{m}$-particle in the toric code insulator. However, the $\mathbb{Z}_4$ topological order is anomalous, in the sense that it can only be realized on the surface of a 3D system but not in a strictly 2D system provided that \textit{both} TRS and translation symmetry are preserved. This is a consequence of the bulk topology. Breaking either symmetry allows for a 2D topological order. In fact, by breaking TRS, one can condense the $d^2$-anyon (which requires back-scattering of helical modes), and reduce the $\mathbb{Z}_4$ state to the $\mathbb{Z}_2$ toric code. As long as the translation symmetry is still preserved, the surface topological order is then identical to our symmetry-enriched toric code insulator. This again suggests that 2D side surfaces of 3D weak TIs are possible venues for realizing the toric code insulator. 

A second route towards a weak TSC makes use of an array of nanowires, with each wire realizing a 1D TSC (i.e. Kitaev chain). Two broad experimental platforms for assembling such 1D quantum wires have attracted much attention: proximitized quantum wires with strong spin-orbit coupling (e.g. InAs and InSb) \cite{Lutchyn2010, Oreg2010, Lutchyn2018}, and magnetic atomic chains on superconductors \cite{Choy:2011p195442,Martin:2012p144505,Gangadharaiah:2011p036801,Nadj-Perge:2013p020407,Klinovaja:2013p186805,Braunecker:2013p147202,Vazifeh:2013p206802}. Both classes of systems have provided promising experimental evidence for the existence of Majorana end states~\cite{Mourik2012,Nadj-Perge:2014p602}. Platforms based on the combination of helical magnetism and superconductivity may provide the most fruitful route to exploring 2D architectures and weak TSC phases. Aiming to generalize the setup based on 1D magnetic atomic chains, initial theoretical proposals have explored the possibility of exploiting modulated magnetic phases in 2D systems to generate effective spin-orbit coupling and local Zeeman splittings. In principle, this allows for the realization of effectively spinless gapped $p+ip$ superconductors~\cite{Nakosai:2013p180503} as well as nodal 2D superconductors~\cite{Sedlmayr:2015p115415}. The latter may be viewed as an intermediate nodal phase separating a trivial SC from a weak TSC. It is important to note, however, that all these proposals and platforms or realizing engineered 1D or 2D TSCs rely on proximity coupled (conventional $s$-wave) superconductivity. This presents a important challenge, since our construction of the toric code insulator relies on intrinsic superconductivity with a fluctuating phase. Hence, even if realized, disordering the resulting weak TSC may be challenging.

A third class of systems relevant to 2D weak TSCs is characterized by quasi-1D Fermi surfaces associated with orbital degrees of freedom. The anisotropic nature of non-$s$-shell orbitals can give rise to directionally anisotropic and strongly quasi-1D hopping. When electrons on such quasi-1D Fermi surfaces form unconventional pairing states of $p$-wave type with a full pairing gap, the resulting phase can realize a weak TSC. This scenario has indeed been proposed for Sr$_2$RuO$_4$ \cite{Hughes:2014p235123}, which has quasi-1D Fermi surfaces coming from $d_{xz,yz}$ orbitals, in addition to a 2D Fermi surface sheet derived from a $d_{xy}$ orbital. Although the nature of the superconducting order parameter of Sr$_2$RuO$_4$ remains an unsettled question, at least partially, one compelling proposal for the pairing state assumes dominant pairing of the quasi-1D Fermi surfaces \cite{Raghu:2010p136401}, which would imply nontrivial weak indices \cite{Hughes:2014p235123}. 

When taking the spin degree of freedom of electrons into account and assuming absence of spin-orbit coupling in the normal state, such pairing leads to a state which is equivalent to two copies of a weak TSC, one for a spin sector~\cite{Hughes:2014p235123}. As argued before, each copy can be disordered into a $\mathbb{Z}_2$ toric code insulator. Hence, in the limit of vanishing spin-orbit coupling, a possible fate of such systems is a $\mathbb{Z}_2 \times \mathbb{Z}_2$ fractionalized insulator with translation symmetry-enrichment. Being a fully gapped topological order, this exotic state is expected to survive when SOC is switched-on adiabatically. We thus believe that Sr$_2$RuO$_4$, and systems alike, could provide a route to exploring generalizations of the toric code insulator introduced in this work.

\subsection{Model realization}

To further aid the identification of experimental platforms, we now introduce a square lattice tight-binding model for the weak TSC. Without pairing, this square lattice model describes the transition between a trivial and a (inversion symmetric) weak TI in 2D, which necessarily occurs via a 2D Dirac semi-metallic phase. Including pairing terms of the Kitaev type gives rise to a weak TSC. Notably, as we will argue below, the 2D model presented here enables an interesting connection with recent proposals for achieving fractionalized phases in 3D weak TSCs, particularly for those realized by gapping 3D Weyl semimetals~\cite{Wang:2020p096603,Thakurathi:2020p235168}. This suggests that the toric code insulator fits into a broader and more general framework for studying and realizing fractionalization in strongly interacting weak topological phases. 

\begin{figure}[t]
   \includegraphics[width=\columnwidth]{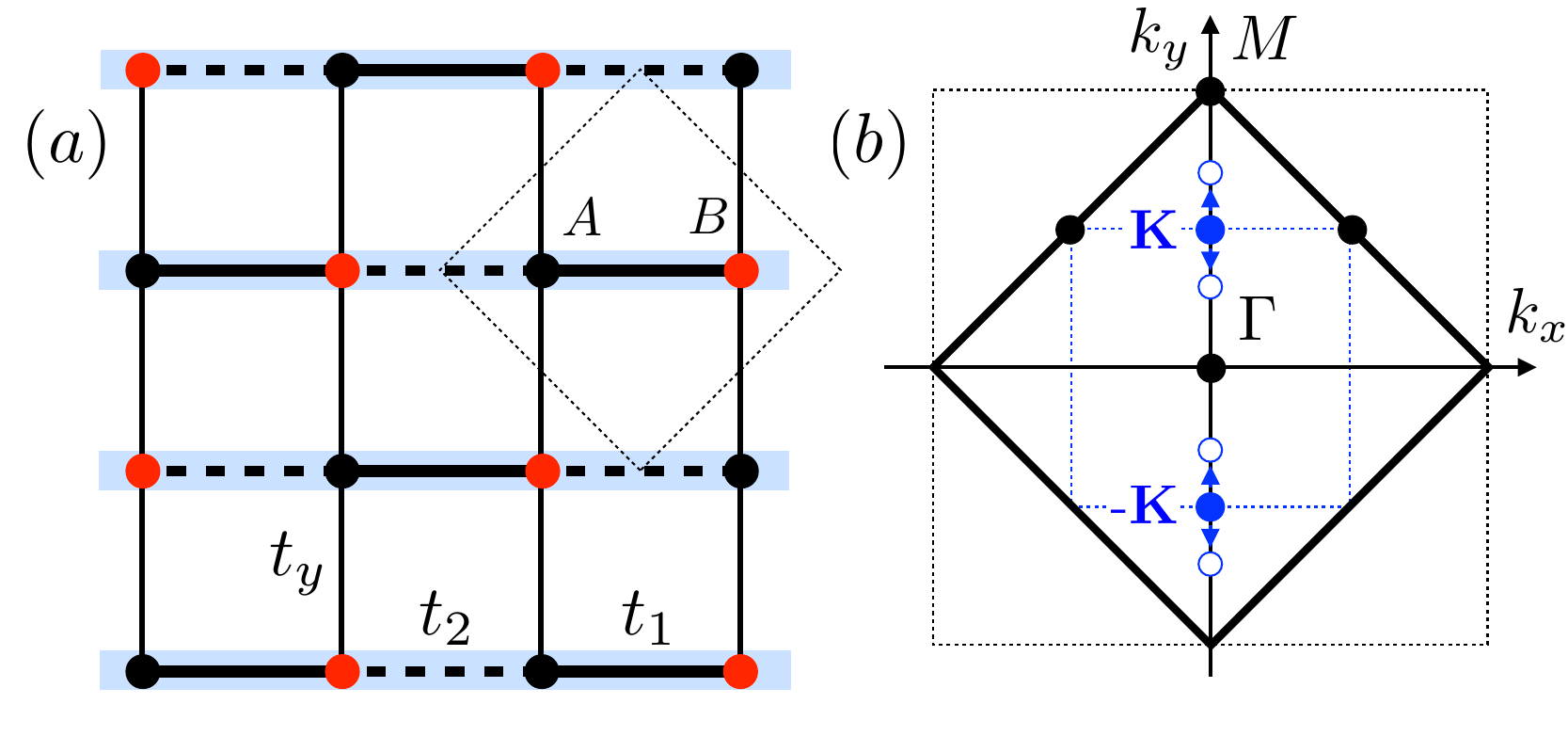}\centering
  \caption{\small{Idealized square lattice model for a weak TSC in two dimensions. (a) The square lattice model of Eq.~\eqref{H_k} is defined by staggered alternating hoppings in the $x$ direction, $t_1$ (intra-cell hopping) and $t_2$ (inter-cell hopping), giving rise to a two-site unit cell (dashed square). The horizontal rows, indicated in light blue, are interpreted as 1D wires (see text). (b) Brillouin zone of the two-site square lattice model. The location of the Dirac nodes which occur when $t_x = (t_1+t_2)/2=0$ are shown by bold blue dots. The splitting of the Dirac nodes into Bogoliubov-Dirac nodes is indicated by open blue dots.}}
  \label{square_model}
\end{figure}

Here we consider a tight-binding model of spinless electrons on the square lattice with alternating hoppings $t_{1,2}$ in the $x$ direction, as depicted in Fig.~\ref{square_model}. The alternating hoppings give rise to a two-site unit cell and we label the two sublattices as $A$ and $B$. The two-component electron operator $c^\dagger_{\bk}$ and the Hamiltonian $H_0$ are given by
\begin{equation}
c^\dagger_\bk = (c^\dagger_{\bk A},c^\dagger_{\bk B}), \qquad H_0 = \sum_\bk c^\dagger_\bk h_\bk c_\bk, \label{H_square}
\end{equation}
The Hamiltonian matrix can be expressed as $h_\bk = \gamma^{\phantom{*}}_\bk \sigma_+ +  \gamma^*_\bk \sigma_-$, where $\sigma_\pm = (\sigma_x\pm i\sigma_y)/2$ are Pauli matrices and $\gamma^{\phantom{*}}_\bk$ is given by
\begin{equation}
\gamma^{\phantom{*}}_\bk  = 2 t_y \cos k_y + t_1e^{ik_x} + t_2 e^{-ik_x}. \label{hoppings}
\end{equation}
The lattice constant has been set to unity. It is useful to decompose the hopping in the $x$ direction into a uniform ($t_x$) and a staggered ($t_0 > 0$) component by writing $t_{1,2} = t_x \mp t_0$. With this parametrization, $h_\bk$ takes the form
\begin{equation}
h_\bk =\varepsilon_\bk \sigma_x +2t_0 \sin k_x\sigma_y, \label{H_k}
\end{equation}
where $\varepsilon_\bk = 2(t_x \cos k_x  + t_y \cos k_y) $ describes an anisotropic square lattice dispersion. The Hamiltonian has two important symmetries: time-reversal ($\mathcal{T}$) and inversion ($\mathcal{P}$) symmetry. $\mathcal{T}$-symmetry is defined by $h^*_\bk =h_{-\bk}$, and $\mathcal{P}$-symmetry by $\sigma_x h_\bk \sigma_x =h_{-\bk}$.

To understand the electronic phases described by this model it is instructive to consider two special cases, which correspond to setting either $t_x$ or $t_y$ to zero. First, consider the case $t_y=0$. We may then view the system as a collection of decoupled 1D wires in the $x$ direction, each realizing a Su-Schrieffer-Heeger (SSH) chain with Hamiltonian $h(k_x) =  2 t_x \cos k_x\sigma_x +2t_0 \sin k_x\sigma_y$. In each chain two distinct insulating phases are possible: a trivial phase and a topological phase. The topological phase is characterized by a fractional charge polarization and is protected by inversion symmetry~\cite{Hughes:2011p245132,Ramamurthy:2015p085105}. This motivates an interpretation of the square lattice model as a collection of 1D wires of spinless electrons stacked and coupled in the $y$ direction. 

A second special case is realized when $t_x=0$ and corresponds to an ordinary square lattice model with anisotropic nearest neighbor hopping in the $x$ and $y$ directions and a flux $\Phi=\pi$ piercing through each square plaquette. As a result of the $\pi$-flux, the spectrum exhibits linear Dirac nodes located at $\pm \bK$, where $\bK = (0,\pi/2)$ (see Fig.~\ref{square_model}b), which shows that the square lattice model describes a 2D Dirac semimetal when $t_x=0$. Nonzero $t_x$ shifts the location of the Dirac nodes along the $k_y$ axis, either towards $\Gamma$ ($t_x < 0$) or towards $M$ ($t_x > 0$) of the Brillouin zone (BZ), see Fig.~\ref{square_model}, where the two nodes can annihilate and gap the spectrum. Importantly, the two insulating phases resulting from Dirac node annihilation, either at $\Gamma$ or at $M$, are topologically distinct. Gapping at $\Gamma$ corresponds to a trivial insulator, while gapping at $M$ corresponds to a \emph{weak} topological insulator in 2D protected by inversion symmetry~\cite{Ramamurthy:2015p085105}. The weak TI phase is adiabatically connected to a stack of decoupled inversion symmetric SSH chains in the topological phase (i.e. with filling anomaly), whereas the trivial insulator is adiabatically connected to a stack of 1D trivial insulators. This further emphasizes the interpretation of the square lattice model as a wire model. Note that here the presence of inversion symmetry is crucial for a robust topological distinction of the electronic phase in 1D, and therefore also for the weak TI.

This analysis demonstrates that the square lattice model of Eq.~\eqref{H_k} describes the transition between a trivial insulator and a (inversion symmetric) weak TI, which necessarily occurs via an intermediate 2D Dirac semimetal~\cite{Ramamurthy:2015p085105}. The case $t_x=0$, with Dirac points at $\pm \bK$, can be viewed as the exact ``half way point'' of this transition. Our next step is to introduce pairing terms and show that this can promote the square lattice model to a weak TSC. We first consider BCS-type zero momentum pairing and introduce a mean field pairing Hamiltonian of the form
\begin{equation}
H_\Delta = \frac12 \sum_\bk  (\Delta_\bk)_{\alpha\beta}  c^{\dagger }_{\bk \alpha}c^{\dagger }_{-\bk \beta} + \text{H.c.}. \label{BCS}
\end{equation}
Here, $\alpha,\beta$ label the sublattice degree of freedom and $\Delta_\bk$ is the pairing potential, which satisfies $\Delta^T_{-\bk}=-\Delta_\bk$. We focus on pairing along the wires in the $x$ direction and consider a nearest neighbor pairing potential of the form $\Delta_\bk = \Delta_0 \cos k_x \sigma_y $. Setting the chemical potential $\mu$ to zero, we obtain a quasiparticle spectrum with four branches given by the equation $(E^\pm_\bk)^2 = (2t_0 \sin k_x)^2 + (\varepsilon_\bk \pm  |\Delta_0| \cos k_x)^2$.
To understand the quasiparticle spectrum, it is useful to consider the case $t_x=0$ and examine the existence of nodes. Given the solution for the quasiparticle spectrum, the condition for vanishing pairing gap is $ 2t_y\cos k_y \pm  |\Delta_0|  =0$, which implies that nonzero (and small) pairing $|\Delta_0|$ splits each Dirac node into two Bogoliubov-Dirac nodes. The Bogoliubov-Dirac nodes move in opposite directions along the $k_y$ axis as $|\Delta_0|$ is changed, as is indicated in Fig.~\ref{square_model}(b). When $|\Delta_0|=2 |t_y|$ the four Bogoliubov-Dirac nodes merge and annihilate in pairs at both $\Gamma$ and $M$, and the resulting fully gapped superconductor realizes a weak TSC in 2D. This may be understood by taking $t_y$ to zero, which does not close the quasiparticle gap, but yields decoupled 1D superconducting wires, each realizing a Kitaev chain~\cite{KitaevChain}. This shows that for sufficiently strong BCS pairing of the Kitaev type (i.e., nearest neighbor pairing along the wires), the square lattice model describes a weak TSC composed of coupled 1D Kitaev chains coupled in the $y$ direction. 

Whereas zero momentum BCS pairing does not immediately lead to a full gap (but instead gives rise to nodal points), finite momentum FFLO pairing directly gaps out the quasiparticle spectrum, and similarly leads to a weak TSC phase. To describe this, we consider finite momentum pairing of the form  
\begin{equation}
H_\Delta = \frac12 \sum_\bk  (\Delta_\bk)_{\alpha\beta}  c^{\dagger }_{\bk \alpha}c^{\dagger }_{-\bk+\bQ \beta} + \text{H.c.}, \label{FFLO}
\end{equation}
where $\bQ=(0,\pi)$ is the wave vector connecting $\Gamma$ to $M$, see Fig.~\ref{square_model}(b). The wave vector $\bQ$ also connects the Dirac points located at $\bK$ and $-\bK$ when $t_x=0$, and we therefore set $t_x$ to zero at first. Including the FFLO pairing term, we obtain the Hamiltonian matrix
\begin{equation}
\mathcal H_\bk = \begin{pmatrix} h_\bk & \Delta_\bk \\  \Delta^\dagger_\bk & -h^T_{-\bk+\bQ} \end{pmatrix},
\end{equation}
where $\Delta_\bk$ satisfies $\Delta^T_{-\bk+\bQ} = -\Delta_{\bk}$ due to Fermi statistics. As a result, the admissible nearest neighbor pair potential takes the same form as in the case of BCS pairing. As before, we focus on pairing of the form $\Delta_\bk = \Delta_0 \cos k_x \sigma_y$ and determine the quasiparticle spectrum; we find two degenerate branches given by
\begin{equation}
E^2_\bk =(2t_0 \sin k_x)^2 + (2t_y\cos k_y)^2 +|\Delta_0|^2 \cos^2 k_x.
\end{equation}
This describes a fully gapped superconductor for arbitrary strength of pairing $\Delta_0$ and the resulting gapped phase is a weak TSC. Note that due to the commensurability of $\bQ$, this FFLO pairing state does not break translation symmetry. Indeed, density modulations of finite momentum $\bQ$ pairing have wave vector $2\bQ$, which is equal to a reciprocal lattice vector.

It is interesting and enlightening to connect this analysis to recent work on Weyl semimetals and weak topological superconductors in 3D. Conceptually, $T$-breaking Weyl semimetals can be viewed as gapless phases describing the topological transition between a trivial insulator and a 3D quantum Hall insulator, i.e., a weak topological phase equivalent to stacks of 2D Chern insulators. Recent work explored different ways of introducing a full pairing gap in this type of Weyl semimetals, with the aim of realizing a 3D weak TSC equivalent to a stack of 2D chiral $p+ip$ superconductors~\cite{Wang:2020p096603,Thakurathi:2020p235168}. In particular, it was shown that commensurate finite momentum FFLO pairing leads to a direct pairing gap~\cite{Wang:2020p096603,Cho:2012p214514,Bednik:2015p035153}, whereas zero momentum BCS pairing requires sufficiently strong pairing in order to induce a merging and pairwise annihilation of Bogoliubov-Weyl nodes at high symmetry points~\cite{Thakurathi:2020p235168,Li:2018p067003}. In both cases, the resulting gapped superconductor realizes a 3D weak chiral TSC with one chiral Majorana mode per stacked layer on any side surface. The case of the $T$-breaking Weyl semimetal is the direct 3D analog of the square-lattice model we have introduced and analyzed in this section. 

\subsection{Impurity effects}
To conclude our discussion on experimental realizations, let us address the fate of toric code insulator in the presence of impurities, which locally breaks the translation symmetry. This begs the question of whether the advertised notion of ``translation symmetry-enrichment" still apply to realistic materials. As argued for other topological phases protected by crystalline symmetries, such as weak topological insulators and topological crystalline insulators, their classification and physical features indeed survive random disorders provided that the symmetry is respected \textit{on average} \cite{Mong2012, Stern2012, FuKane2012}. Here we argue that the same rationale applies to the toric code insulator.

Two main aspects of the effect of disorder can be addressed: one about the gapless edge and one about the gapped bulk. For the edge, the symmetry-enriched toric code insulator has gapless Majorana modes, which is identical to the edge of an undimerized weak TSC. In the latter context, the stability of edge modes in the presence of random disorder has been established by Morimoto and Furusaki \cite{weakTSCstability2014}: the edge Hamiltonian has a unique mass term of the dimerization type, and when the disorder-average of the mass term vanishes, the edge state remains critical and gapless. As for the bulk, the effect of disorder is even simpler to understand: the gapped topological order, including the anyon spectrum and braiding statistics, cannot be altered by any weak disorder incapable of closing the bulk gap. In particular, the fermion-parity-switching effect would still constrain the motion of a single vison to hop across two wires at a time. As long as the notion of ``even-link vs odd-link" remains, there are always \textit{two} types of visons ($\mathbf{m}_e$ and $\mathbf{m}_o$), and they still have mutual semionic braiding statistics. Last but not least, $\mathbf{m}_e$ and $\mathbf{m}_o$ are still related by a discrete translation across a wire (though it is not necessarily a ``symmetry" anymore). In short, the above reasoning
provides a realistic scenario for realizing the toric code insulator in the proposed experimental platforms.

\section{\label{conclusion} Discussion and conclusion}

In this paper we have introduced and analyzed a coupled wire construction for a translation symmetry-enriched $\mathbb{Z}_2$ topological order, which is termed the toric code insulator. Our construction is based on the vortex-condensation approach, for which the weak TSC in 2D (i.e., a gapped superconductor equivalent to an array of 1D TSCs) serves as the conceptual starting point. The nature of the 2D weak TSC naturally suggests a coupled wire model description. A key feature of the weak TSC is the fractional Josephson effect, which gives rise to a distinction between two types of vortices: vortices on even and odd links. In this way, the fractional Josephson effect is inextricably linked to the translational properties of weak TSC, as well as to the implementation of translation symmetry in the toric code insulator phase. Starting from the weak TSC, the latter is the result of proliferating double-vortices.

To describe double-vortex condensation, we have introduced a three-fluid model for the coupled wires. In addition to a Luttinger liquid of charge-$e$ fermions and a Luttinger liquid of charge-$2e$ Cooper pairs in each wire, this model consists of a third fluid, which arises from a $\mathbb{Z}_2$ gauge field defined on the link. The $\mathbb{Z}_2$ gauge field is a consequence of gauging the superconducting phase-shift symmetry and achieves a decoupling of the charge and neutral sectors of the two Luttinger liquids. In this sense, the three-fluid model--particularly the introduction of the $\mathbb{Z}_2$ gauge field---bears resemblance to previously considered slave-particle approaches and parton constructions \cite{SenthilFisher}. Importantly, the magnetic flux of the $\mathbb{Z}_2$ gauge field corresponds to a single vortex in the phase of the Cooper-pair fluid.  

Our central result is the analysis of the three-fluid model and the demonstration that it leads to a 2D gapped insulator with topological order---a new phase of matter which we refer to as the toric code insulator. The topological order is of the $\mathbb{Z}_2$ type akin to the toric code, with the following anyons: charge-$e$ chargon ($\mathbf{e}$), neutral fermion ($\mathbf{f}$), and two types of visons ($\mathbf{m}$). An electron $\psi_e$ can fractionalize into a chargon and a neutral fermion, thus exhibiting a charge-statistics separation. Moreover, there is a translation symmetry-enrichment manifested in the toric code insulator: translation symmetry relates two types of visons, $\mathbf{m}_e$ (on even-links) and $\mathbf{m}_o$ (on odd-links), which have a semionic mutual braiding statistics. One can thus associate $\mathbf{m}_e$ to the $e$-anyon, and $\mathbf{m}_o$ to the $m$-anyon in the original language of Kitaev's toric code. Particularly, the $e \leftrightarrow m$ anyonic permutation (which is a ``symmetry" for the topological order as all the fusion and braiding properties are preserved) has been realized here as a real-space translation. The toric code insulator is thus a tensor product of a physical electron and a toric code topological order enriched by translation symmetry: $\psi_e \otimes \{\mathbf{m}_e, \mathbf{m}_o, \mathbf{f}\}$. The complete topological data is summarized in Table. \ref{spectrum}.

Our work suggests that strongly-interacting weak SPT phases are promising platforms for realizing SET phases enriched by translation symmetry. In particular, neither the Hamiltonian nor the ground state breaks translation symmetry, yet translation has a non-trivial effect as \textit{permuting} anyons (here $\mathbf{m}_e \leftrightarrow \mathbf{m}_o$). In other words, the pattern of anyonic excitations \textit{breaks} the symmetry of Hamiltonian. This phenomenon has been termed ``weak symmetry breaking'' \cite{Kitaev2006exactly}, and is an intriguing consequence of symmetry-enrichment in topological order. There are celebrated spin lattice models, such as Wen's plaquette model and Kitaev's honeycomb model \cite{Wenplaquette2003, Kitaev2006exactly}, which feature this effect. Recently, Rao and Sodemann have proposed to understand the weak breaking of translation symmetry in these models as a consequence of the emergent spinons forming a weak TSC, which leads to a mobility constraint for $m$-particles in the toric code \cite{Rao2020}. This is essentially the same pattern of symmetry-enrichment as realized in the toric code insulator introduced here. There is, however, a crucial difference: our system is built out of itinerant electrons, instead of localized magnetic moments on a lattice. This perspective has enabled us to consider several material realizations, which include the surface of 3D weak TI, 2D array of nanowires or magnetic adatom chains, and correlated materials with quasi-1D Fermi surfaces. 

Furthermore, our work has an interesting connection with recent proposals for strongly correlated fractionalized phases in 3D Weyl semimetals~\cite{Wang:2020p096603,Thakurathi:2020p235168}. In particular, Ref.~\onlinecite{Wang:2020p096603} explored the possibility of realizing a 3D fractional quantum Hall effect in Weyl semimetals by disordering a 3D weak TSC. The 3D weak TSC can be realized by pairing fermions within each Weyl node and is equivalent to a stack of 2D chiral $p+ip$ Read-Green superconductors. This is analogous to the 2D weak TSC, which can be understood as a stack of 1D Kitaev TSCs. As demonstrated in the previous section, the 2D weak TSC is intimately related to topological semimetals in 2D. Our work therefore shows that gapping and disordering topological semimetals by proliferating defects should be considered a general route towards realizing novel types of topological order, both in 2D and 3D. The two canonical examples of topological semimetals in 2D and 3D suggest possible generalizations to other types of topological semimetals, for which spatial symmetries play a prominent role.\\

\begin{acknowledgments}
This work is in part supported by the Croucher Scholarship for Doctoral Study from the Croucher Foundation (P.M.T.) and a Simons Investigator grant from the Simons Foundation (C.L.K.). 
\end{acknowledgments}

\appendix
\renewcommand{\theequation}{\thesection.\arabic{equation}}
\section{\label{AppA} The Gauss's law constraint}

In the main text, we have treated the $y$-component of gauge field $A_{y,\ell}$ as the density variable $\Theta_\ell$ of a Luttinger liquid, and the conjugate electric field is then $E_{y,\ell} = \partial_x \Phi_\ell$. To facilitate this Luttinger liquid representation, we have to adopt a gauge-fixing condition: $A_x=0$, but then the operator $e^{iE_x/2}$ which tunnels a $\pi$-flux (of $B = \partial_x A_y -\Delta_y A_x$) is \textit{incompatible} with this gauge. We thus look for a re-writing of this operator using the Gauss's law constraint, which we now derive.

Due to the discreteness of the gauge fields, one should not attempt to perform an infinitesimal variation of $A_t$ in hope of obtaining an equation of motion that represents the Gauss's law. Instead, the Gauss's law is obtained by integrating out $A_t$ in a discretized fashion. This procedure can be carried out straightforwardly by viewing the $\mathbb{Z}_2$ gauge theory (coupled to a compact $\varphi_2$) as a $\mathbb{Z}$ gauge theory (coupled to a non-compact $\varphi_2$). In this case, the ``basic" configuration of $A_t$ takes the following form:
\begin{equation}\label{At}
A_t(j,x,t) = \pi n_j H(x-x_0) \delta(t-t_0),
\end{equation}
with $n_j \in \mathbb{Z}$ and $H(x)$ being the Heaviside step function. Here $x_0$ and $t_0$ label a reference space-time position where the gauge field changes its discrete value. A more general form of $A_t$ results from the superposition of these ``basic" configurations. Now the functional integral $\int \mathcal{D}A_t$ can be replaced by a discrete sum $\sum_{n_j}$. Let us only consider the part of the action associated to $A_t$, which arises from minimal coupling, and integrate it out:
\begin{widetext}
\begin{equation}
\begin{split}
&\mathcal{Z} \sim \int \mathcal{D}A_t \exp \Big\{\sum_j \int_{x,t}  \frac{i}{2\pi}\big[\partial_x{\theta_{2,j}}(\partial_t\varphi_{2,j} -2 A_{t,j}) -E_{x,j}\partial_x A_{t,j}-E_{y,\ell}\Delta_yA_{t,j}\big]\Big\} \\ 
& \sim \prod_j \sum_{n_j\in \mathbb{Z}} \exp \Big\{i n_j \big[ \theta_{2,j} -\frac{1}{2}E_{x,j} -\frac{1}{2}(\Delta_y \Phi)_j\big]\Big\} \;\;= \;\; \prod_j \sum_{m_j \in \mathbb{Z}}4\pi \delta\Big(2\theta_{2,j}-E_{x,j} -(\Delta_y\Phi)_j -4\pi m_j\Big)
\end{split}
\end{equation}
\end{widetext}
The second to last expression is obtained upon substituting Eq. (\ref{At}). The last equality is obtained from the Poisson summation, and the delta function imposes the Gauss's law constraint as quoted in Eq. (\ref{Gausslaw}). In the above derivation, we tentatively choose a field configuration as in Eq. (\ref{At}), so the obtained constraint holds at $(x_0, t_0)$ for every wire $j$. Since we do not have to specify $(x_0, t_0)$, and indeed the complete functional integral should counts all possible configurations, the Gauss's law constraint holds in general. The $\pi$-flux hopping operator can thus be written as in Eq. (\ref{twist}):
\begin{equation}
e^{\frac{i}{2}E_{x,j}} = e^{i[(\theta_\rho-\theta_\sigma)_j-\frac{1}{2}(\Delta_y\Phi)_j]}.
\end{equation}
Notice the right-hand side is consistent with the $A_x=0$ gauge, so this serves as a physical operator in our Luttinger liquid formalism. This is used in the Sec. \ref{sec3.1} to study the motion of visons in the toric code insulator, which reveals a pattern of translation symmetry-enrichment. 

\section{\label{AppB} Counting GSD in the wire model}
The Wilson-loop argument in Sec. \ref{sec3.2} suggests there is a topologically-protected lower bound to the ground state degeneracy (GSD) of the symmetry-enriched toric code insulator on torus ($T^2$). Due to the interplay between topology and translation symmetry, the torus GSD depends on the parity of the number of wires, which is referred to as the even-odd effect as summarized in Eq. (\ref{GSD}). Here we count the GSD on $T^2$ explicitly using the three-fluid wire model developed in Sec. \ref{sec2.3}.

Ground states are determined by the interaction potentials in Eq. (\ref{3fluidH}):
\begin{equation}
\mathcal{H}_{\text{int}} = \sum^L_{j=1} (u\cos{\varphi_{\sigma,j}}+w\cos{2\theta_{\rho,j}}+h\cos{2\Theta_{\ell}}).
\end{equation}
with $\ell \equiv j+1/2$. We consider the system with $L$ wires on a torus by imposing periodic boundary conditions in both the $x$-(along the wire) and the $y$-(along the stacking of wires) directions, so wire $j = L+1$ is identified with wire $j=1$. In the strong-coupling limit, ground states are gapped and characterized by the condensed values of $\varphi_{\sigma,j}$, $\theta_{\rho,j}$ and $\Theta_{\ell}$, which are respectively pinned to the bottom of the cosine potentials. Hence, $\varphi_{\sigma,j} \in 2\pi \mathbb{Z}$, $\theta_{\rho,j} \in \pi\mathbb{Z}$ and $\Theta_{\ell} \in \pi\mathbb{Z}$. Notice that the bosonic fields are compact:
\begin{subequations}\label{compa}
\begin{align}
\varphi_{\sigma,j} &\equiv \varphi_{\sigma,j} + 4\pi; \\
\theta_{\rho, j} &\equiv \theta_{\rho,j} +2\pi; \\
 \Theta_{\ell} &\equiv \Theta_{\ell} +2\pi,
\end{align}
\end{subequations}
which imply that there are \textit{at most} $2^{3L}$ distinct ground states.

On top of the above compactifications, there are other redundancies in the definition of the bosonic fields, which lead to identification in the $2^{3L}$ states we just naively counted. The redundancies in the bosonic fields are revealed by local operators that create kinks in them. Considering the local electron operator $\psi_e$, in Eq. (\ref{electronop}), we see that the charge and neutral sectors have an intertwined identification:
\begin{equation}\label{IA}
I_{A,j}:\; (\varphi_{\sigma,j}, \theta_{\rho,j}) \equiv (\varphi_{\sigma,j} +2\pi, \theta_{\rho,j}+\pi).
\end{equation}
To be precise, the above should also be accompanied by a $\pi$-shift in $\theta_{\sigma,j}$, but for simplicity we would leave it implicit. After all, what matter to the determination of ground states are $\{\varphi_\sigma, \theta_\rho, \Theta \}$, hence we will only focus on these variables. Next, there is an intertwined identification of fields on neighboring wires, as revealed by the action of $T_\ell (\mathbf{f})$ introduced in Eq. (\ref{Tf}). We have
\begin{equation}\label{IB}
I_{B,\ell}:\; (\varphi_{\sigma,j}, \varphi_{\sigma,j+1}) \equiv (\varphi_{\sigma,j}+2\pi,\varphi_{\sigma,j+1}+2\pi).
\end{equation}\\
Finally, by virtue of the $\mathbb{Z}_2$ gauge redundancy, 
\begin{equation}\label{IC}
I_{C,j}: (\varphi_{\sigma,j}, \Theta_{\ell},\Theta_{\ell-1}) \equiv (\varphi_{\sigma,j}+2\pi, \Theta_{\ell} +\pi,\Theta_{\ell-1}+\pi),
\end{equation}
which is also revealed by the action of the twist operator $e^{iE_{x,j}/2}$ in Eq. (\ref{twist}). The above three types of identification fully capture the redundancies in the definition of fields. For each wire $j$ there are identifications $I_{A,j}$ and $I_{C,j}$ that can be performed to relate equivalent ground-state configurations. Similarly, for each link $\ell$ there is an identification $I_{B,\ell}$ that can be used to relate equivalent configurations. Starting from the $2^{3L}$ counting, each \textit{independent} identification reduces the number of distinct ground states by a factor of 2. 

Crucially, not all redundancies introduced above are independent. Due to the torus geometry, 
\begin{equation}
\prod_{j=1}^L I_{B,j+\frac{1}{2}} = 1.
\end{equation}
This is because applying $I_{B,\ell}$ on every link leads to a $4\pi$-shift of $\varphi_\sigma$ for every wire, which have been considered already in Eq. (\ref{compa}). Hence there are only $(L-1)$ independent redundancies associated to Eq. (\ref{IB}). 

Now it comes the even-odd effect. If $L$ is even, there is one more relation (thus one less redundancy):
\begin{equation}
\prod_{j=1}^L I_{C,j} = \prod_{i=1}^{L/2} I_{B,2i+\frac{1}{2}},
\end{equation}
as the net effect of the left-hand side (after modding out the compactification of $\Theta_\ell$) is to induce a $2\pi$-shift of $\varphi_\sigma$ on every wire, which is equivalent to applying Eq. (\ref{IB}) on every alternating link. Everything considered, the GSD on $T^2$ in the three-fluid wire model is thus
\begin{equation}
\text{GSD} = 
\begin{cases}
2^{3L} \times 2^{-(3L-2)} = 4,\;\text{for even }L; \\
2^{3L} \times 2^{-(3L-1)} = 2,\;\text{for odd }L.
\end{cases}
\end{equation}

\bibliographystyle{apsrev4-1.bst}

\bibliography{TCInsulator}

\end{document}